# Extension of measurement range in OCDR based on double-modulation scheme

Mana Sakamoto,[1,2] Tomoya Miyamae,[1] Kohei Noda,[1,3,4]
Kentaro Nakamura,[3] Heeyoung Lee,[2] and Yosuke Mizuno[1]

[1] *Faculty of Engineering, Yokohama National University, 79-5 Tokiwadai, Hodogaya-ku, Yokohama 240-8501, Japan*
[2] *Graduate School of Engineering and Science, Shibaura Institute of Technology, 3-7-5 Toyosu, Koto-ku, Tokyo 135-8548, Japan*
[3] *Institute of Innovative Research, Tokyo Institute of Technology, 4259, Nagatsuta-cho, Midori-ku, Yokohama 226-8503, Japan*
[4] *Graduate School of Engineering, The University of Tokyo, 7-3-1 Hongo, Bunkyo-ku, Tokyo 113-8656, Japan*
*Author e-mail addresses: sakamoto-mana-fh@ynu.jp, mizuno-yosuke-rg@ynu.ac.jp*

**Abstract:** We extend the measurement range of optical correlation-domain reflectometry (OCDR) by modulating the laser output frequency at two frequencies, while preserving spatial resolution. We demonstrate distributed reflectivity sensing with a ten-fold extended measurement range.

## 1. Introduction

In recent years, the widespread adoption of optical fiber communication networks has enabled high-speed and high-capacity data transmission. However, the reliability of these networks is often hindered by faults in the optical fiber cables, leading to an increasing demand for diagnostic tools that can identify the location and reflectivity of connection failures and other reflection points [1]. Distributed reflectivity sensors, which can measure the position and reflectivity of faults on optical fibers, have attracted attention in response to this demand.

Optical time-domain reflectometers (OTDRs) [2–4] and optical frequency-domain reflectometers (OFDRs) [5–7] are widely used for this purpose, with the former having a longer measurement range and the latter having a higher spatial resolution. However, OTDRs struggle to achieve real-time operation with a high signal-to-noise ratio, while OFDRs require expensive narrow-linewidth lasers. In order to overcome these limitations, an optical correlation-domain reflectometer (OCDR) based on the synthesis of optical coherence functions (SOCF) was proposed [8–10]. The OCDR has the advantages of random accessibility (allowing fast measurements at any position on the measured fiber) and real-time operation, and can operate with relatively inexpensive lasers.

In the case of the OCDR using Brillouin scattering (BOCDR), it is known that spatial resolution and measurement range are in a trade-off relationship [11]. Meanwhile, it has been believed that the spatial resolution of the OCDR is not affected by the measurement range. However, recent theoretical and experimental studies have revealed that the spatial resolution of the OCDR is also in a trade-off relationship with the measurement range [12]. Therefore, methods to extend the measurement range while maintaining the spatial resolution of the OCDR are required.

In this study, we first demonstrate that the dual-modulation method proposed for the BOCDR [13] can be used to extend the measurement range of the OCDR while maintaining spatial resolution.

## 2. Principles

The basic principle of SOCF-OCDR is illustrated in Fig. 1. A sinusoidal modulation is applied to a laser, generating correlation peaks (measurement points) in the fiber under test (FUT). By controlling the modulation frequency, the position of the correlation peak is swept along the FUT, enabling the measurement of the reflection distribution [8,9]. While the measurement range (correlation peak spacing) is inversely proportional to the modulation frequency, it has been believed that the spatial resolution is only inversely proportional to the modulation amplitude. However, recent studies have revealed that the spatial resolution is also inversely proportional to the modulation frequency, in addition to the modulation amplitude [12]. Thus, the spatial resolution and measurement range of OCDR are in a trade-off relationship.

The double-modulation scheme is used to overcome this trade-off in the case of BOCDR [13]. In this method, a sinusoidal modulation is applied to a laser by mixing two different frequencies, $f_0$ and $m f_0$, where $f_0$ is the fundamental frequency and $m$ is an integer. As shown in Fig. 2, correlation peaks generated by modulation at $m f_0$, which are not multiples of m, are suppressed by modulation at $f_0$. Furthermore, by setting the modulation amplitude at $f_0$ relatively low, $f_0$ no longer affects the spatial resolution. As a result, the double-modulation scheme enables the compatibility of the spatial resolution determined by $m f_0$ and the measurement range determined by $f_0$.

## 3. Experiments

### 3.1 Methods

The experimental setup for OCDR with double modulation is shown in Fig. 3. We constructed it based on the simplified OCDR [14] that eliminated an acousto-optic modulator. A delay fiber of 5 km was inserted into the reference path. Polarization-dependent fluctuations were suppressed by using a polarization scrambler. The interference signals between the reference and reflected lights were converted into electrical signals and observed using an oscilloscope with a zero-span function of an electrical spectrum analyzer to monitor the power as a function of time at 1.0 MHz for 2 $f_0$, 1.5 MHz for 3 $f_0$, 2.0 MHz for 4 $f_0$, and 3.4 MHz for 10 $f_0$. We performed double modulation of $f_0$ and 2 $f_0$, $f_0$ and 3 $f_0$, $f_0$ and 4 $f_0$, and $f_0$ and 10 $f_0$, and scanned the frequencies from $f_0$ at approximately 381.0 kHz to 404.0 kHz, 2 $f_0$ at approximately 762.0 kHz to 808.0 kHz, 3 $f_0$ at approximately 1.14 MHz to 1.20 MHz, 4 $f_0$ at approximately 1.52 MHz to 1.62 MHz, and 10 $f_0$ at approximately 3.81 MHz to 4.04 MHz (using different channels of the same waveform generator). The modulation amplitudes were set to 0.8 GHz for $f_0$ and 3.0 GHz for 2 $f_0$, 3 $f_0$, 4 $f_0$, and 10 $f_0$. For comparison, we also measured the reflectivity distribution with single modulation of only 2 $f_0$, 3 $f_0$, 4 $f_0$, and 10 $f_0$.

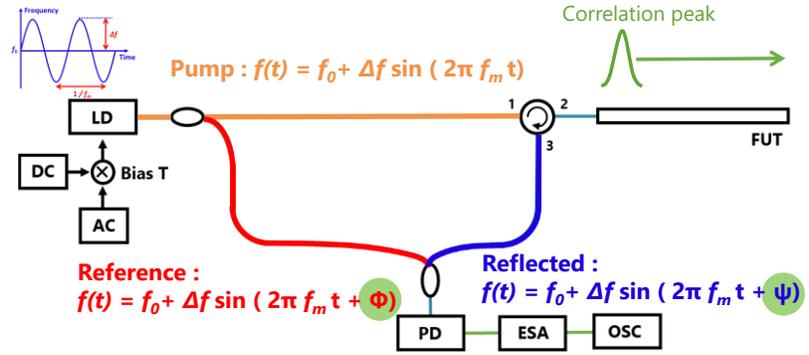

Fig. 1. Operating principle of optical correlation-domain reflectometry (OCDR).

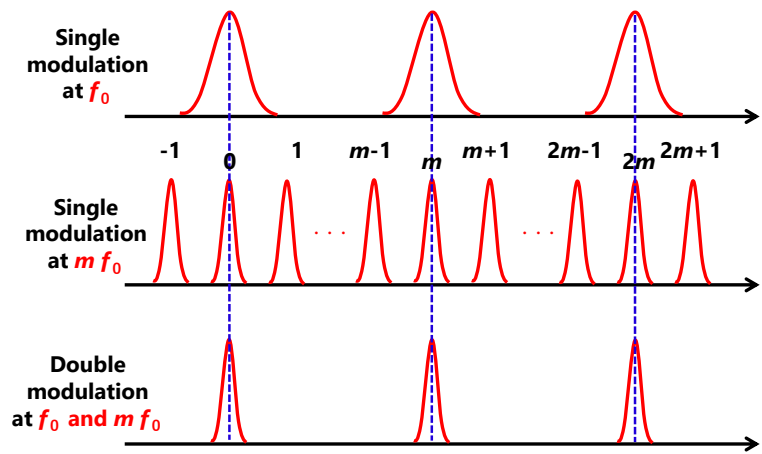

Fig. 2. Operating principle of double-modulation scheme.

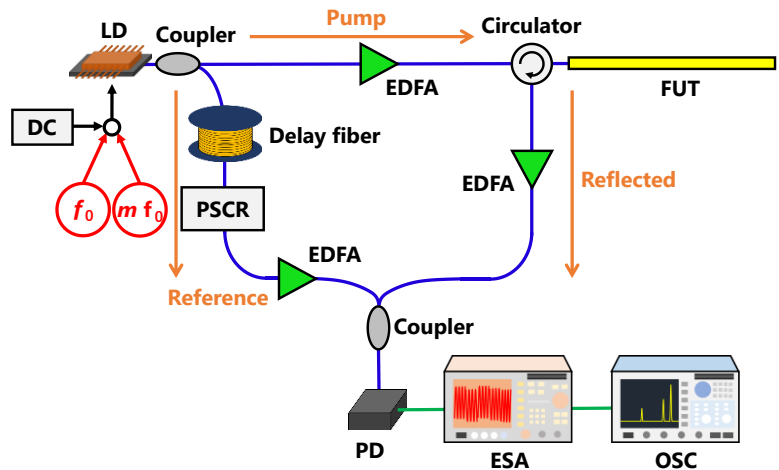

Fig. 3. Experimental setup of OCDR based on double-modulation scheme. EDFA: erbium-doped fiber amplifier, ESA: electrical spectrum analyzer, FUT: fiber under test, LD: laser diode, OSC: oscilloscope, PD: photodetector, PSCR: polarization scrambler.

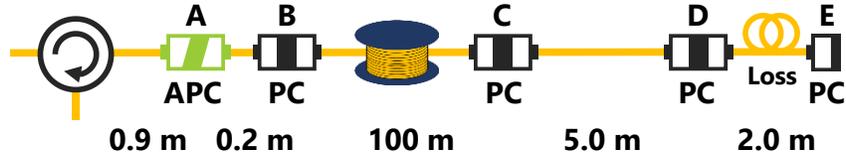

Fig. 4. Schematic structure of the FUT.

The configuration of the FUT is shown in Fig. 4. Relatively strong reflections were generated at connectors B, C, D, and E (bend loss was applied before the open end E). We measured the reflectivity distribution from the open end to a position 30 m away for double modulation of 2 $f_0$, 80 m away for 3 $f_0$, 60 m away for 4 $f_0$, and 84 m away for 10 $f_0$ from the circulator (assuming the detection of three reflection points from connectors C, D, and E).

### 3.2 Demonstration of double-modulation OCDR

First, we show the results of the reflectivity distribution measurement with single modulation of 2 $f_0$ in Fig. 5(a). The correct peak at connectors C, D, and E was detected, but a dummy peak caused by connector B appeared at around 35 m, which was not the correct result compared to the actual reflection point, because multiple correlation peaks exist inside the FUT with single modulation (the measurement range was insufficient compared to the length of the FUT). Next, we show the results of double modulation of $f_0$ and 2 $f_0$ in Fig. 5(b). The peak caused by connector B was suppressed, and the correct measurement was obtained. The results on the cases where $m$ = 3, 4 are omitted in this manuscript due to the space limitation.

Finally, the results with single modulation at 10 $f_0$ are shown in Fig. 6(a). As with the measurements with single modulation at 2 $f_0$, 3 $f_0$, and 4 $f_0$, the original peaks at connectors C, D, and E were correctly detected, but a dummy peak due to connector B appeared at approximately 86 m, resulting in an incorrect measurement compared to the actual position. When double modulation was applied at $f_0$ and 10 $f_0$ (Fig. 6(b)), the peak due to connector B was similarly suppressed, enabling accurate measurement. In this case, the phenomenon of splitting of each reflection peak was observed, which should be further studied in the future.

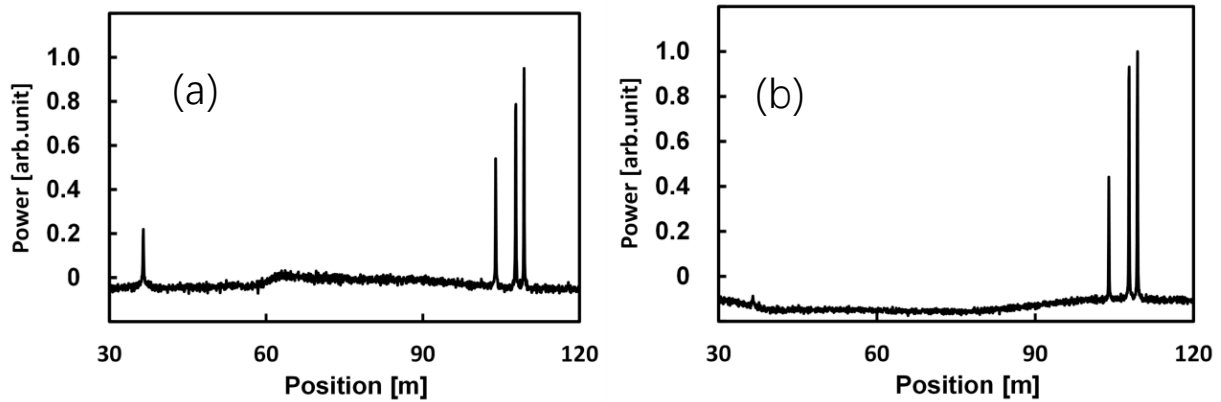

Fig. 5. Reflected power distributions with modulation at (a) 2$f_0$ only and (b) $f_0$ and 2$f_0$.

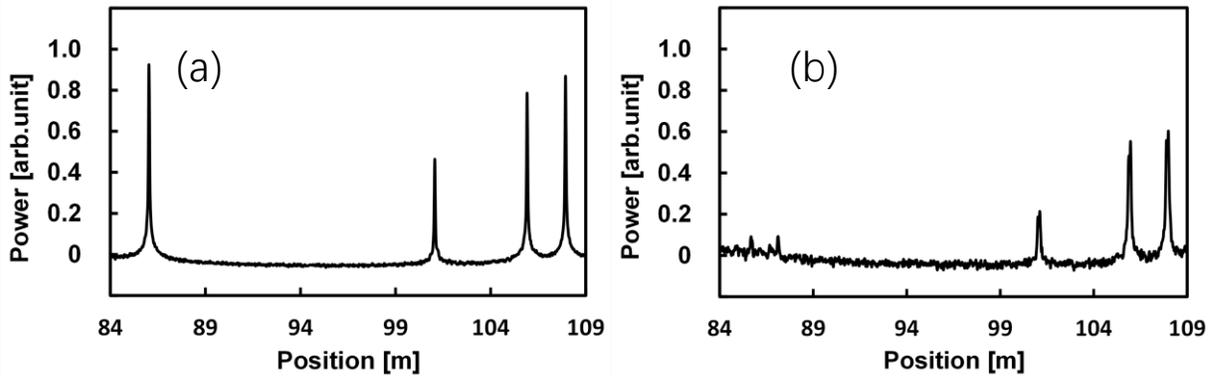

Fig. 6. Reflected power distributions with modulation at (a) $10 f_0$ only and (b) $f_0$ and $10 f_0$.

## 4. Conclusions

We demonstrated the effectiveness of the double-modulation scheme for OCDR. We showed that it enables an extension of the measurement range by a factor of up to 10 while maintaining the spatial resolution. In future work, we will investigate the shape of the beat spectra in OCDR based on the double-modulation scheme. This scheme will also be beneficial in enhancing the performance of a correlation-domain LiDAR (light detection and ranging) system [15] in the future.

**Acknowledgements**

This work was supported in part by the Japan Society for the Promotion of Science (JSPS) KAKENHI under Grant Nos. 21H04555, 22K14272, and 20J22160.